\documentclass[preprint2]{aastex}
\begin{document}

\title{Enhancing Our Knowledge of Northern Cepheids through Photometric Monitoring}

\author{David G. Turner, Daniel J. Majaess, David J. Lane}
\affil{Saint Mary's University, Halifax, Nova Scotia, Canada}
\affil{The Abbey Ridge Observatory, Stillwater Lake, Nova Scotia, Canada}
\email{turner@ap.smu.ca}

\author{L. Szabados}
\affil{Konkoly Observatory, Budapest, Hungary}

\author{V.~V. Kovtyukh, I.~A. Usenko}
\affil{Odessa National University, Odessa, Ukraine}

\author{L. N. Berdnikov}
\affil{Sternberg Astronomical Institute, Moscow, Russian Federation}

\begin{abstract}
A selection of known and newly-discovered northern hemisphere Cepheids and related objects are being monitored regularly through CCD observations at the automated Abbey Ridge Observatory, near Halifax, and photoelectric photometry from the Saint Mary's University Burke-Gaffney Observatory. Included is Polaris, which is displaying unusual fluctuations in its growing light amplitude, and a short-period, double-mode Cepheid, HDE 344787, with an amplitude smaller than that of Polaris, along with a selection of other classical Cepheids in need of additional observations. The observations are being used to establish basic parameters for the Cepheids, for application to the Galactic calibration of the Cepheid period-luminosity relation as well as studies of Galactic structure.
\end{abstract}
\keywords{astronomical data bases: surveys---stars: variables: Cepheids---stars: variables: other}

\section{Introduction}

The retirement of many experienced photometrists worldwide and the closure of many small telescopes used in their studies has led to a decline in recent photometric monitoring of Cepheid variables. At the same time, such monitoring has grown in importance with the realization that the study of Cepheid period changes can yield vital information on the location of individual variables within the Cepheid instability strip \citep{1}.

In response to the decline in available photometric data for northern hemisphere Cepheids, a program of regular monitoring of bright Cepheids was initiated in 1996 using a CCD imager on the 0.4-m telescope of the Burke-Gaffney Observatory (BGO) at Saint Mary's University \citep{2,3}, and continued intermittently as weather and funding permitted. Photometric monitoring of bright Cepheids ({\it e.g.}, Polaris) was initiated at the BGO in 2003 \citep{4} with the acquisition of an Optec SSP-3 solid state photometer, and the CCD program was extended to fainter objects in winter and summer 2006, including unstudied suspected Cepheid variables. In September 2006 the CCD monitoring program was moved to the 0.3-m Celestron telescope of the automated Abbey Ridge Observatory \citep{11,5}, and presently collects a few hours of observation of Cepheids and other suspected variables on every available clear night --- amounting to several hundred hours of monitoring each year. Extended runs have also been made for special objects, {\it e.g.}, primary minimum for eclipsing systems and small amplitude Cepheids. Some of the more interesting results from such studies are presented here. 

\begin{figure}[ht]
\includegraphics[width=0.45\textwidth]{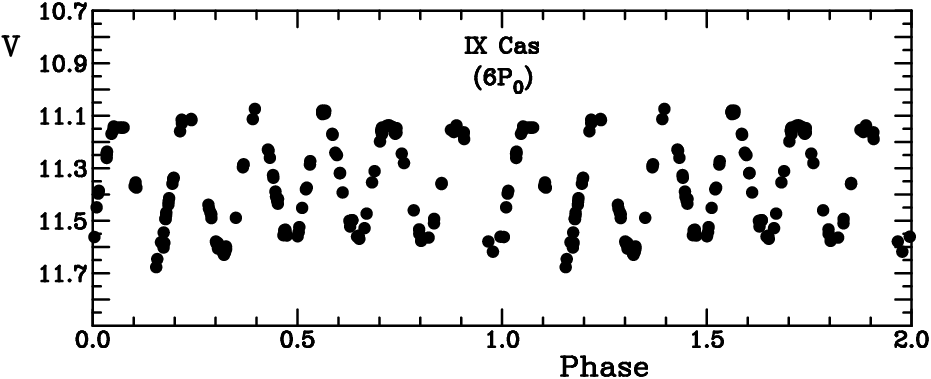}
\caption{\label{ixcas} \small{ARO observations of IX Cas phased to $6 \times P_0$.}}
\end{figure}

\section{Results for Individual Cepheids}
\subsection{IX Cas}
This binary Cepheid of pulsation period $P \simeq 9^{\rm d}$ is categorized as Type II on the basis of metallicity \citep{7}, but its rapid period changes also suggest a first-crossing object \citep{2}. Regular monitoring from the ARO to establish its rate of period change revealed curious temporal changes in mean brightness of the star that, while perhaps typical of Type II variables \citep{8}, may be linked to the star's unusual binarity: 3--5 $M_{\odot}$ companion in a close $110^{\rm d}.49$ orbit \citep{6}. The light curve for six times the present $9^{\rm d}.157$ pulsation period (Fig.~\ref{ixcas}), {\it i.e.}, $\sim \frac{1}{2} P_{\rm orb}$, displays sinusoidal variations similar to nearside heating in close binaries, and averaging of the same data (to remove the pulsational variations) phased to the $110^{\rm d}.49$ orbital period suggests the possibility of eclipses in the system.  Further work is needed.

\begin{figure}[h]
\includegraphics[width=0.43\textwidth]{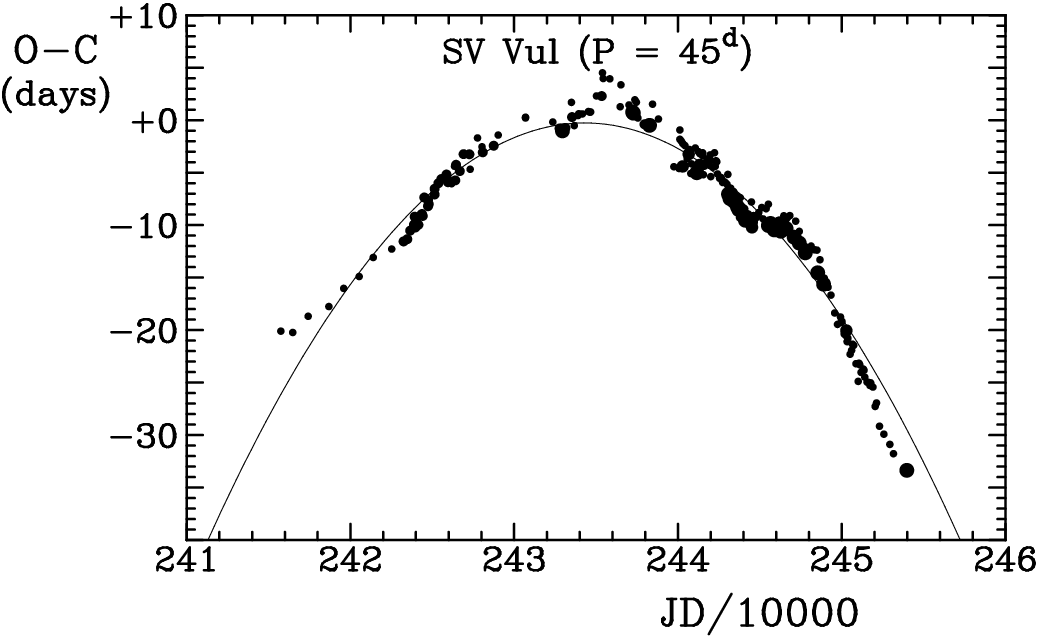}
\includegraphics[width=0.43\textwidth]{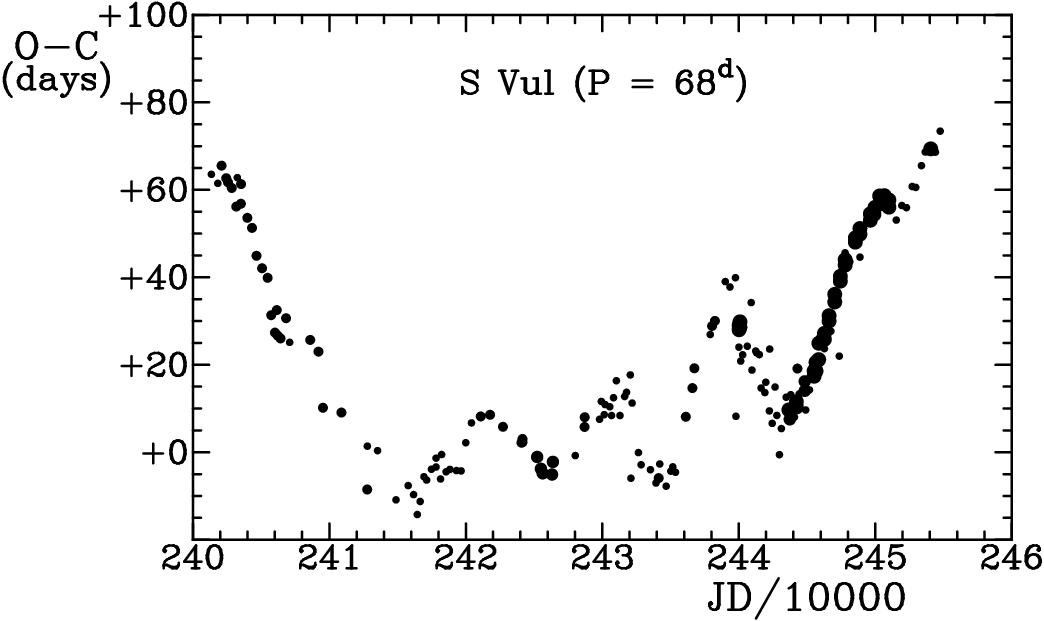}
\caption{\label{ocdiagrams} \small{The O--C variations of SV Vul (upper) and S Vul (lower), with the size of individual data points proportional to the weight assigned to the value.}}
\end{figure}

\subsection{Cepheids with Exotic O--C Diagrams}
This category includes long-period Cepheids like SV Vul and S Vul that exhibit random fluctuations in pulsation period superposed upon the evolutionary trends in their O--C diagrams (Fig.~\ref{ocdiagrams}), in the case of SV Vul suggesting a recent increase in its rate of blueward evolution \citep{10}, and recognized binaries like X Cyg (not illustrated). The O--C data for SV Vul and S Vul include analyses of AAVSO and ARO observations for the stars.

\begin{figure}[h]
\begin{center}
\includegraphics[width=0.33\textwidth]{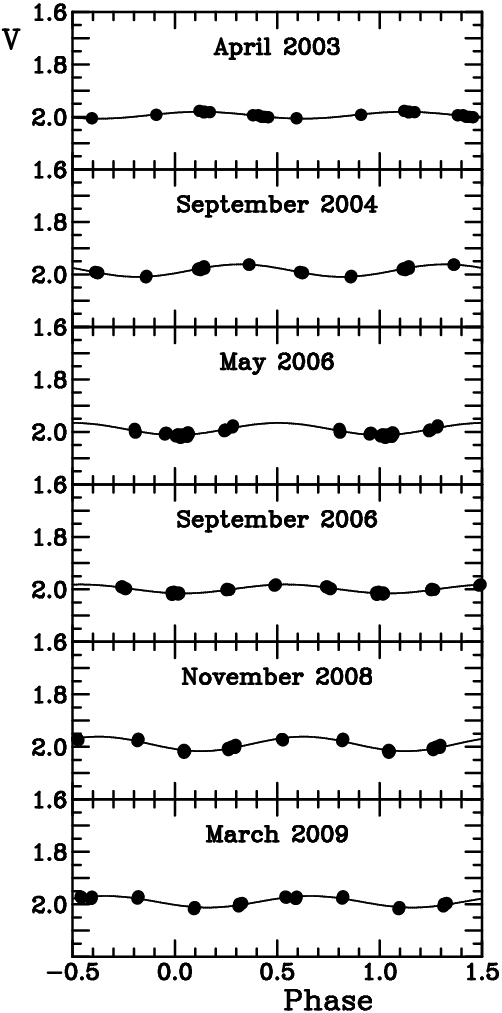}
\caption{\label{polaris} \small{Light curves of Polaris obtained from observations at the Burke-Gaffney Observatory.}}
\end{center}
\end{figure}
\subsection{Polaris}
Regular photoelectric monitoring of Polaris is done at the BGO during nights of good seeing in periods of the year when photometric weather occurs regularly: typically mid-winter, spring, and autumn. The star is observed differentially relative to HD 5914 (spectral type A3 V, $V = 5.86$), and the magnitude scale is adjusted for small air mass differences between it and Polaris. Typical sets of observations from 2003 to 2009 are shown in Fig.~\ref{polaris}, from which one can detect the gradual changes in phase resulting from the regular period increase of Polaris, as well as its gradually increasing pulsation amplitude.

\begin{figure}[h]
\includegraphics[width=0.45\textwidth]{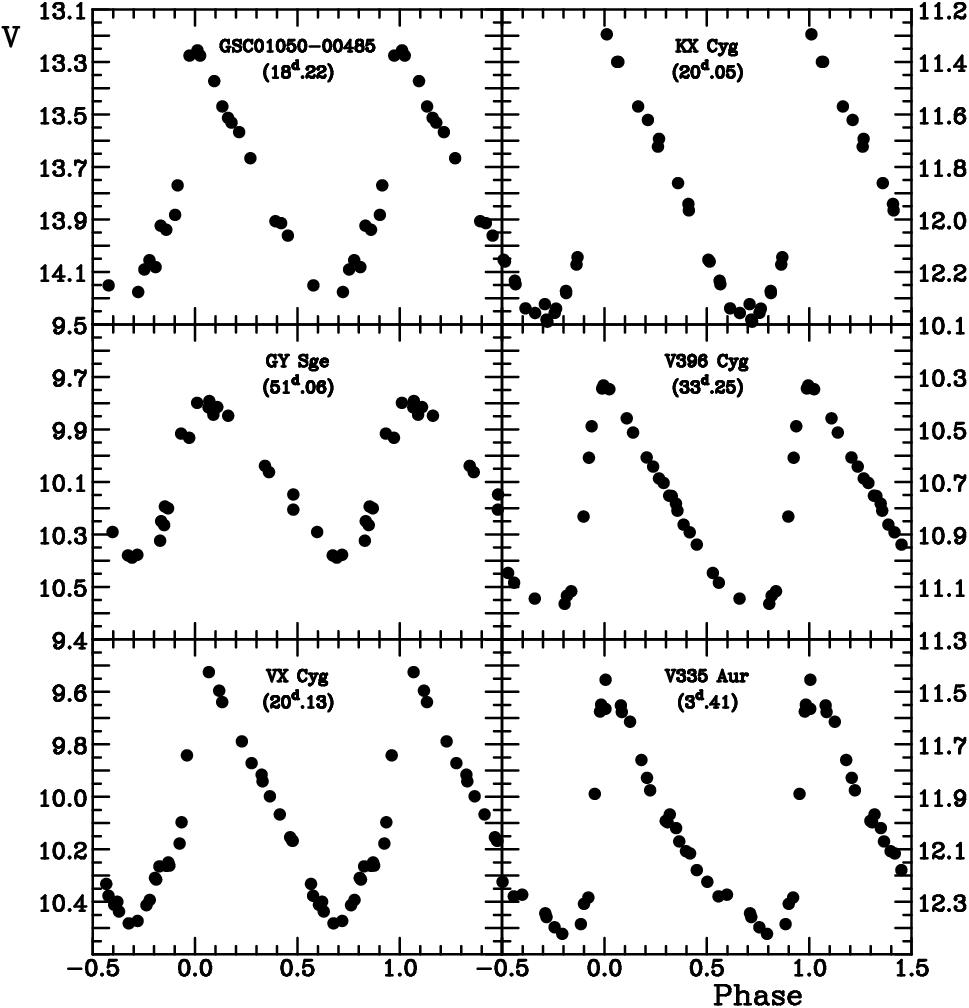}
\caption{\label{ncepheids} \small{Observations of neglected Cepheids from the Abbey Ridge Observatory.}}
\end{figure}
\subsection{Neglected Cepheids}
The list includes a selection of long-period Cepheids lying in the first Galactic quadrant that are important for mapping spiral structure in the Milky Way, newly discovered Cepheids from the NSV catalogue with periods established from the observations, and a selection of short-period Cepheids with peculiar O--C changes. Sample light curves depicted in Fig.~\ref{ncepheids} are not necessarily tied to proper zero-points, and are intended merely to display the schematic features of the observed light curves in the {\it V}-band.

\subsection{HDE 344787}
\begin{figure}[h]
\begin{center}
\includegraphics[width=0.33\textwidth]{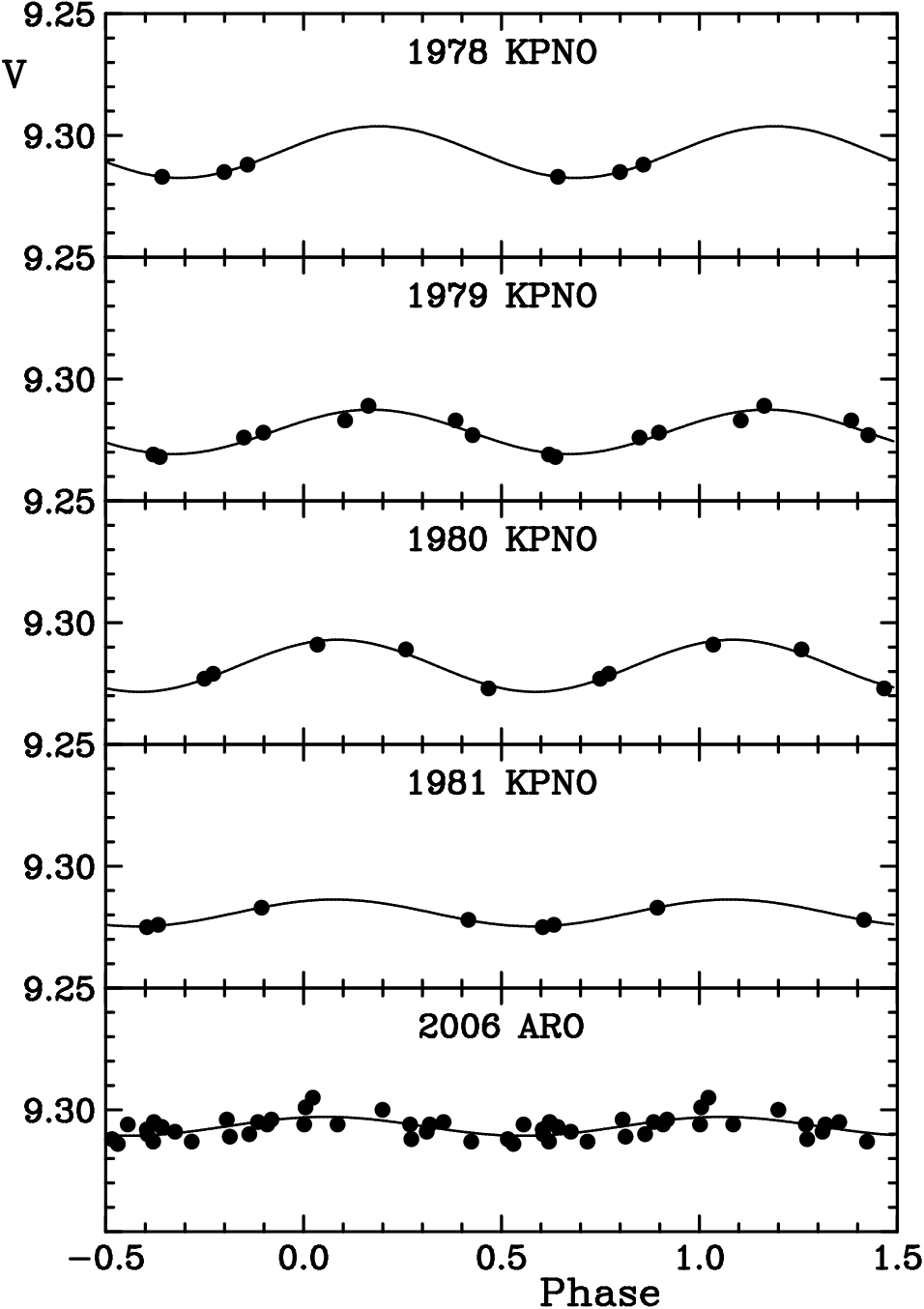}
\includegraphics[width=0.4\textwidth]{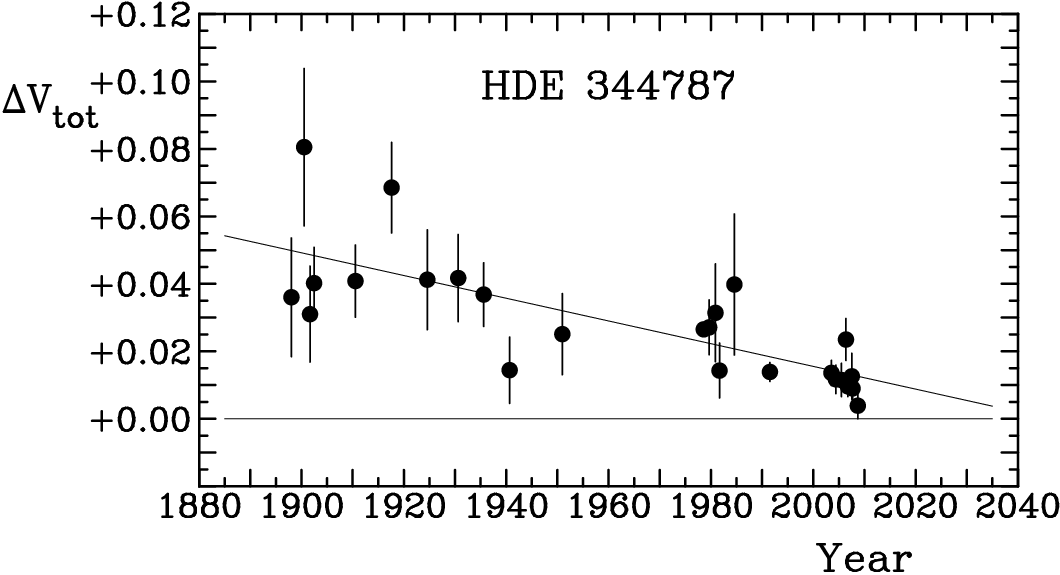}
\caption{\label{hde344787} \small{Photoelectric and CCD light curves for HDE 344787 from observations at Kitt Peak National Observatory and the Abbey Ridge Observatory (upper).  Evidence for a declining light amplitude in the Cepheid HDE 344787 (lower).}}
\end{center}
\end{figure}

The F9 Ib supergiant BD$+22^{\degr}3786$ was suspected to be a possible Cepheid three decades ago, but has only recently been confirmed as such \citep{9}. The star's pulsations have been tracked from the 1890s using the photographic plate archives of Harvard College Observatory, along with observations from Kitt Peak from two decades ago (Fig.~\ref{hde344787}, top). Recent monitoring from the ARO confirms the variability of the star, which has a smaller amplitude than Polaris, but also reveals a rapidly dropping light amplitude (Fig.~\ref{hde344787}, bottom) that may signal cessation of variability by 2045. Typical uncertainties in CCD photometry obtained from automated images exposed on multiple nights of variable sky quality, with the inevitable air mass and instrumental uncertainties, are large enough ($\pm0^{\rm m}.02$ to $\pm0^{\rm m}.03$) to make continued monitoring of the Cepheid challenging.

\section{Summary}
Program objects being monitored from our two Halifax sites include a variety of other variable stars, in addition to the large sample of Cepheids, of which only a partial listing is presented here. A separate program of regular monitoring of southern hemisphere Cepheids is operated by Berdnikov, using facilities in Chile and elsewhere. This paper illustrates some of the more unusual results obtained with the available facilities.

\end{document}